\documentclass[manuscript]{aastex631}
\usepackage{amsmath}

\shorttitle{Blazhko and Non-Blazhko RRLs }
\shortauthors{P. Zong et al.}
\graphicspath{{./}{figures/}}

\begin{document}

\title{Pulsation Properties of Blazhko and Non-Blazhko RRab Stars}

\correspondingauthor{Jian-Ning. Fu; Jie. Su; Bing-Kai Zhang}
\email{jnfu@bnu.edu.cn; sujie@ynao.ac.cn; zhangbk@fynu.edu.cn}
\author{Peng Zong}
\affiliation{Key Laboratory of Functional Materials and Devices for Informatics of Anhui Higher Education Institutes, School of Physics and electronic engineering,Fuyang Normal University, Fuyang 236037, People’s Republic of China}
\affiliation{School of Physics and Astronomy,Beijing Normal University, Beijing, 100875, People's Republic of China}

\author{Jian-Ning Fu}
\affiliation{School of Physics and Astronomy,Beijing Normal University, Beijing, 100875, People's Republic of China}
\affiliation{Institute for Frontiers in Astronomy and Astrophysics, Beijing Normal University, Beijing 102206, People’s Republic of China}

\author{Jie Su}
\affiliation{Yunnan Observatories, Chinese Academy of Sciences, Kunming 650216, People’s Republic of China}
\affiliation{Key Laboratory for the Structure and Evolution of Celestial Objects, Chinese Academy of Sciences, Kunming 650216, People’s Republic of China}
\affiliation{International Centre of Supernovae, Yunnan Key Laboratory, Kunming 650216, People’s Republic of China}

\author{Bing-Kai Zhang}
\affiliation{Department of Physics, Fuyang Normal University, Fuyang 236041, People’s Republic of China}
\affiliation{Key Laboratory of Functional Materials and Devices for Informatics of Anhui Higher Education Institutes, Fuyang Normal University, Fuyang 236037, People’s Republic of China}

\author{Gao-Chao Liu}
\affiliation{Center for Astronomy and Space Sciences, China Three Gorges University, Yichang 443002, People’s Republic of China}

\author{Weichao Sun}
\affiliation{School of Physics and Astronomy,Beijing Normal University, Beijing, 100875, People's Republic of China}
\affiliation{Institute for Frontiers in Astronomy and Astrophysics, Beijing Normal University, Beijing 102206, People’s Republic of China}

\author{Jiaxin Wang}
\affiliation{School of Science Chongqing University of Posts and Telecommunications, Chongqing, 400065, People's Republic of China}

\author{Bo Zhang}
\affiliation{Key Laboratory of Space Astronomy and Technology,National Astronomical Observatories of China, Chinese Academy of Science, Beijing 100012, People’s Republic of China}

\author{Xueying Hu}
\affiliation{School of Physics and Astronomy,Beijing Normal University, Beijing, 100875, People's Republic of China}
\affiliation{Institute for Frontiers in Astronomy and Astrophysics, Beijing Normal University, Beijing 102206, People’s Republic of China}

\author{Zhongrui Bai}
\affiliation{Key Laboratory of Optical Astronomy, National Astronomical Observatories, Chinese Academy of Sciences, Beijing 100012, People’s Republic of China}

\author{Weikai Zong}
\affiliation{School of Physics and Astronomy,Beijing Normal University, Beijing, 100875, People's Republic of China}
\affiliation{Institute for Frontiers in Astronomy and Astrophysics, Beijing Normal University, Beijing 102206, People’s Republic of China}



\begin{abstract}
In this study, we conduct a comparative analysis of the properties of Blazhko and non-Blazhko RRab stars. We identified 1054 non-Blazhko and 785 Blazhko RRab stars in the photometric data observed by K2 mission, which, combined with those 37 stars observed in the original Kepler field, constituted our study sample. Using the Fourier Decomposition method, we calculated the pulsation parameters, including phase differences and amplitude ratios, for these RRab stars, revealing significant discrepancies in the pulsation parameters between Blazhko and non-Blazhko RRab stars. However, distinguishing between Blazhko and Non-Blazhko RRab stars based on Fourier parameters remains challenging due to the significant overlap in their distributions. By cross-matching our sample with the LRS of LAMOST DR12, we identified 147 Blazhko and 111 non-Blazhko RRab stars, which exhibit similar metallicity distributions. Furthermore, cross-matching with Gaia DR3 data yielded 766 Blazhko and 950 non-Blazhko RRab stars, showing differences in color indices but not in absolute magnitudes. Our findings suggested the Blazhko effect is linked to pulsation parameters and colors, rather than metallicities or absolute magnitude. 
\end{abstract}

\section{Introduction}
The Blazhko effect, characterised by periodic phase or amplitude variations in the light curves, remains a mysterious physical phenomenon since its recognition in the radial pulsating RR Lyrae stars (RRLs) over a century ago by \cite{1907AN....175..325B,chadid2010first}. An interesting parameter
of this effect is the Blazhko frequency, which corresponds to a period spanning from a few days to several years. In the frequency spectrum, this phenomenon is identified through the presence of multiplet peaks situated close to the primary pulsation frequency and its associated lower-order harmonics \citep{2023AJ....165..194W}.Previous studies \citep[see, e.g.,][]{2010ApJ...713L.198K, 2012MNRAS.424.3094S, 2016CoKon.105...61K,2022ApJS..258....8M}, drawing on extensive photometric data from large ground-based surveys and space observations, suggest that nearly half of all fundamental-mode RRab stars exhibit Blazhko modulation. Various theoretical models have been proposed to explain the Blazhko effect \citep{2000ASPC..203..299S, 2004AcA....54..363D, 2006ApJ...652..643S, 2011ApJ...731...24B, 2013A&A...554A..46G, 2013arXiv1310.0543S, 2014ApJ...783L..15B}, but none of them have been conclusively confirmed by observations. The phenomenon of period doubling observed in Blazhko stars \citep{2010MNRAS.409.1244S} suggests that resonances between low- and high-order radial modes may play a key role in explaining the Blazhko effect \citep{2011MNRAS.414.1111K,2011ApJ...731...24B}.

Comparison studies between the properties of Blazhko and non-Blazhko RRab stars have been conducted. Earlier works \citep{2003ApJ...598..597A, 2011MNRAS.411.1763J, 2014MNRAS.445.1584S} suggested that Blazhko RRab stars tend to have slightly shorter pulsation periods compared to Non-Blazhko RRab stars, whereas \cite{2003A&A...398..213M} found no significant difference in the pulsation periods between the two groups. Regarding metallicity, \cite{2003A&A...398..213M} suggested that the occurrence rate of the Blazhko effect increases with higher metallicity. However, investigations by \cite{2005AcA....55...59S} and \cite{2014MNRAS.445.1584S} found no correlation between the Blazhko effect and metallicity. \cite{2011MNRAS.411.1763J} conducted a comparative study of Blazhko and Non-Blazhko RRab stars in M5, finding that Blazhko RRab stars generally exhibit larger amplitudes in their light curves, shorter periods, and fainter magnitudes compared to Non-Blazhko RRab stars. \cite{2014MNRAS.445.1584S} suggested that the differences in properties between Blazhko and Non-Blazhko RRab stars, observed in the study by \cite{2011MNRAS.411.1763J}, require further investigation in other stellar systems, as the findings based on a limited sample of stars in M5 may be misleading. Based on photometric data from ASAS and WASP, \cite{2014MNRAS.445.1584S} suggested weak indications that Blazhko stars might prefer slightly lower metallicity and shorter periods, though no convincing evidence was found in the study.
In terms of the colors of the two groups of stars, \cite{2012MNRAS.420.1333A}, using ground-based photometric data, found no difference between Blazhko and Non-Blazhko RRab stars. However, the time span and sampling intervals of the photometric data in their study were insufficient. The absolute magnitudes and colors of the Blazhko and Non-Blazhko RRab stars in the literature \citep{2011MNRAS.411.1763J, 2012MNRAS.420.1333A, 2014MNRAS.445.1584S} were estimated using the equations of \cite{1998A&A...333..571J}, which were derived based on photometric data. As documented by \cite{2017MNRAS.466.2602P}, these studies are based on varying numbers of stars, each with different quantities and qualities of observations, and cover diverse stellar systems that may differ intrinsically. They emphasized that, to obtain reliable statistical characteristics of Blazhko and Non-Blazhko RRab stars, a large, homogeneous sample of stars with high-quality data is essential.

The unprecedented and invaluable photometric data obtained from the $Kepler$ \citep{2010ApJ...713L..79K}, $K2$ missions \citep{2014PASP..126..398H}, and $TESS$ \citep{2015JATIS...1a4003R}, along with the data provided by $Gaia$ \citep{2018A&A...616A...8A} and the homogeneous metallicities [Fe/H] estimated from the Low-Resolution Spectral (LRS) data of LAMOST \citep{2012RAA....12.1243L, 2015RAA....15.1095L}, have greatly advanced the investigation of RR Lyrae stars. The RRLs observed by $Kepler$ have been identified and extensively investigated in the literature \citep[see, for example,][]{2010MNRAS.409.1585B, 2010ApJ...713L.198K, 2011MNRAS.417.1022N, 2011MNRAS.411..878K, 2012MNRAS.424..649G, 2013ApJ...773..181N}. These studies primarily focus on the pulsation properties of either Blazhko or Non-Blazhko RRab stars based on $Kepler$ photometric data. Among them, only \cite{2013ApJ...773..181N} performed a comparative study of Blazhko and Non-Blazhko RRab stars, and they also conducted spectroscopic observations of all the RRLs in the $\sl Kepler$ field. They found that 5 of the 21 Non-Blazhko RRab stars are more metal-rich than [Fe/H]$\sim$ -1.0 dex, while none of the 16 Blazhko RRab stars exceeds this value. Using the first released photometric data from $\sl TESS$, \cite{2022ApJS..258....8M} found that the pulsation periods of the Blazhko stars are shorter than those of the Non-Blazhko stars. \cite{2023ApJ...945...18Z} conducted an analysis of 57 Non-Blazhko RRab stars using $Kepler$/$K2$ photometric data, along with metallicities [Fe/H] estimated from the LRS of the LAMOST-$Kepler$/$K2$ surveys \citep{2015ApJS..220...19D, 2018ApJS..238...30Z, 2020ApJS..251...27W, 2020RAA....20..167F}. In contrast, some investigations either lack metallicity data \citep{2022ApJS..258....8M} or do not include Blazhko RRab stars \citep{2023ApJ...945...18Z}. While the low-resolution spectral (LRS) survey of  LAMOST aims to determine stellar parameters for a diverse range of targets across the northern hemisphere, specifically those with declinations above -10$^{\circ}$ \citep{2012RAA....12.1243L,2015RAA....15.1095L}. The LAMOST-\(Kepler\)/\(K2\) surveys \citep{2015ApJS..220...19D,2018ApJS..238...30Z,2020ApJS..251...27W,2020RAA....20..167F} have provided extensive multi-epoch spectra for numerous \(Kepler/K2\) targets, offering an opportunity to study the differences in metallicities between Blazhko and Non-Blazhko RRab stars. Studies on RRLs using Gaia data, as discussed in the literature \citep[see, e.g.,][]{2024arXiv240612408N,2024A&A...684A.176P,2023ApJ...951..114Z,2022MNRAS.513..788G}, do not focus on comparing the distinct characteristics of Blazhko and Non-Blazhko RRab stars. However, we can calculate the absolute magnitudes of Blazhko and Non-Blazhko RRab stars using Equation (21) provided by \cite{2024A&A...684A.176P}, which was obtained using metallicities derived from Gaia DR3. And Gaia also provides us an opportunity to explore the difference in colours between Blazhko and Non-Blazhko RRab stars.



In this work, we focus on analyzing the pulsation parameters of Blazhko and Non-Blazhko RRab stars using light curves from the \(Kepler\) and \(K2\) mission to identify differences in pulsation parameters between these two types of stars. The study presents the comparison study of the brightness and colours based on the data provided by \(Gaia\) and explores the metallicities differences estimated from the LRS of LAMOST-$Kepler/K2$ surveys between the Blazhko and Non-Blazhko RRab stars. The paper is organized as follows: Section 2 covers data collection and analysis , Section 3 presents the results, and Sections 4 and 5 provide the discussion and conclusion.

\section{Data Collection and Analysis}
\subsection{Sample}
\begin{figure*}
\begin{center}
\includegraphics[width=7.0 in]{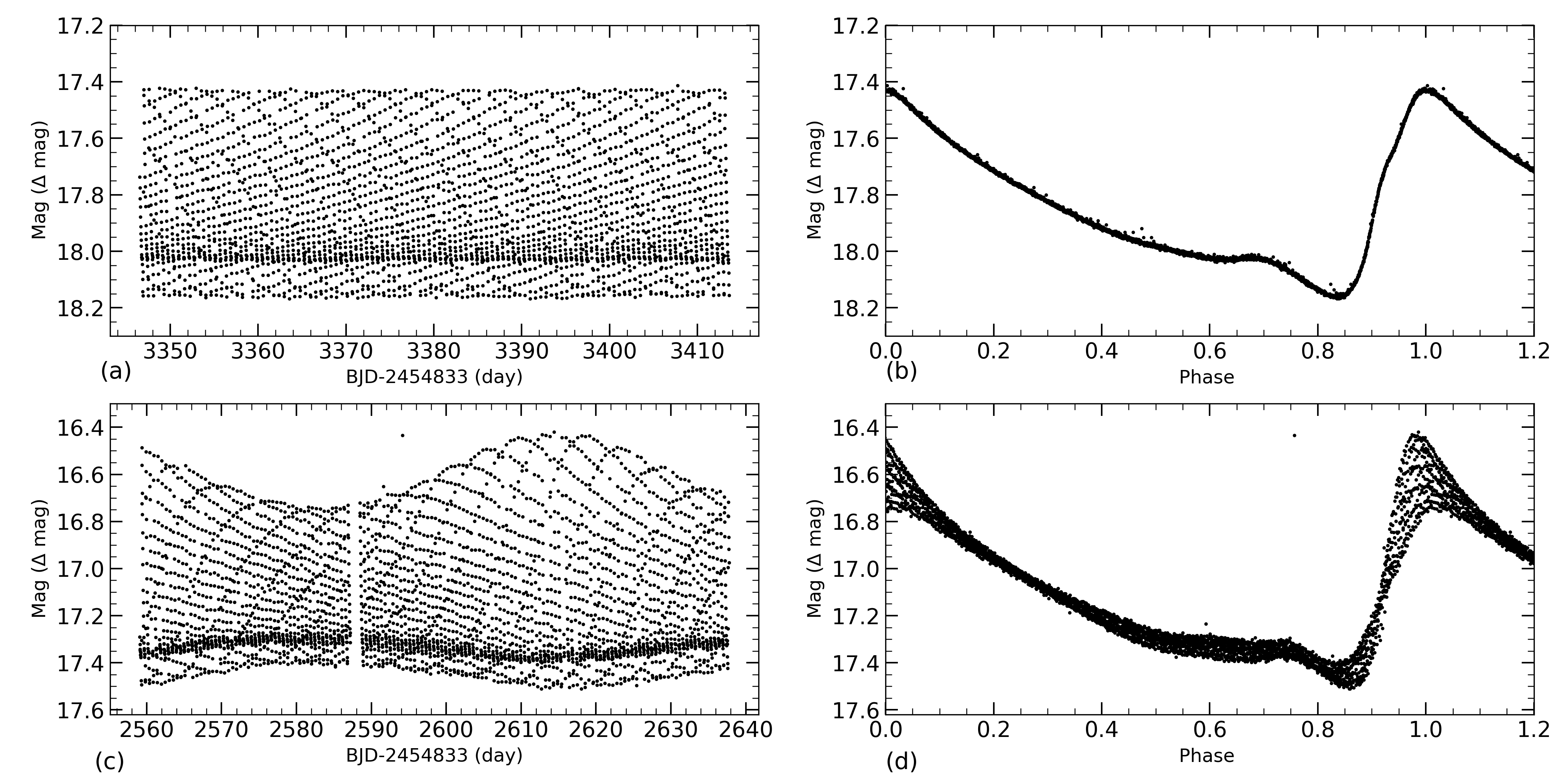}
\caption{The light curves of the Non-Blazhko and Blazhko ab-type RRLs. (a) and (b) show the light curves of the Non-Blazhko ab-type star EPIC 212719863, while (c) and (d) depict the light curves of the Blazhko ab-type star EPIC 220244829. The phased light curves of both stars are folded by their fundamental periods.}
\label{fig:NBL_BL}
\end{center}    
\end{figure*}

The light curves of RRL candidate stars, derived from photometric data observed during Campaigns 0 to 19 (C0-C19) of the \(K2\) mission, were provided by \cite{2022PASP..134a4503B}. According to \cite{2023A&A...677A.177N}, this sample consists of 3917 light curves from the \(K2\) mission, with some candidates observed across multiple campaigns. The total number of unique candidates in the entire sample is 3057, and their sky coordinates are presented in \cite{2023A&A...677A.177N}. In this study, Blazhko and non-Blazhko candidates were identified based on visible light curve modulation or the presence of side peaks in the frequency spectra. The Blazhko effect is typically detected through visual inspection of the light curve, which shows amplitude variations ranging from over 0.5 mag \citep{2012AJ....144...39L} to less than 0.1 mag \citep{2009MNRAS.397..350J}, or even just a few hundredths of a magnitude \citep{2011MNRAS.417..974B, 2023A&A...677A.177N}. In the frequency spectrum, the Blazhko effect manifests as modulation multiplets centered on the pulsation frequency and its harmonics \citep{2014A&A...562A..90S, 2023A&A...677A.177N}. The frequency separation between the side peaks and the pulsation frequency corresponds to the modulation frequency \(f_B\) and its period \(P_B\) \citep{2023A&A...677A.177N}. However, \cite{2023A&A...677A.177N} cautioned that side peaks might arise from observational time span or instrumental artifacts, requiring careful interpretation. Signals were removed as reported in the study of \cite{2022PASP..134a4503B} based on the method provided by in the literature \citep{2015MNRAS.447.2880A,2016MNRAS.456.2260A}. We then scanned for the highest amplitude frequencies and subtracted them using SigSpec \citep{2012CoAst.163....3R} in the 0-30 c/d range, with a signal-to-noise ratio (S/N) $>$ 5.6 \citep{2021ApJ...921...37Z}. We identified 1054 non-Blazhko and 786 Blazhko RRab stars, which, together with the Blazhko and non-Blazhko RRab stars observed in the \(Kepler\) field \citep{2013ApJ...773..181N}, constitute our research sample. Figure~\ref{fig:NBL_BL} presents the a sample of one Non-Blazhko and one Blazhko RRab among those stars derived from the $K2$ mission. 



\subsection{Pulsation Analysis} \label{sec:highlight}
The frequency analysis of the stars was conducted with the software package SigSpec \citep{2012CoAst.163....3R} based on the multi-frequency analysis with Fourier transform and least-square fitting. We followed a standard pre-whitening procedure to analyze the pulsation signals by performing the Discrete Fourier Transform (DFT) to light curves. That is, we select the frequency with the largest amplitude from the DFT spectrum, and then fit the light curve with a sinusoid to correct the frequency value and obtain the amplitude and phase values. After that, the sine signal component is subtracted from the original light curve, we then perform DFT again and repeat the fitting step. However, we should be aware that the pre-whitening process will introduce extra noise. Therefore, in the consecutive pre-whitening steps, we employed the Marquardt-Levenberg algorithm for non-linear fitting. The values of frequency, amplitude, and phase were improved in each step. The purpose of this approach is to minimize the additional noise introduced by the pre-whitening process as much as possible, thereby making the parameters as accurate as possible. 
The Fourier decomposition of the light curves of variable stars, an effective technique for studying their pulsation properties, was first used to analyze the pulsation characteristics of Cepheids \citep{1982ApJ...261..586S} and has since been widely applied to RR Lyrae stars \citep[see, e.g.,][]{1993AJ....106..703S,2011MNRAS.417.1022N,2014MNRAS.445.1584S}. The Fourier decomposition for the Non-Blazhko RRab stars is performed using the following Fourier sine series:
\begin{equation}
\label{eq:1}
m(t) = m_0+\sum \limits_{i=1}^{n} A_i {\rm sin}(2\pi i f_0 (t-t_0)+\phi_{i}),
\end{equation}
where \( n \) represents the number of fitted orders, \( f_0 \) is the main frequency, \( t \) is the observation time (Barycentric Julian Date: BJD - 2454833.0), and \( t_0 \) is the time of the first maximum. The mean magnitude \( m_0 \), amplitude \( A_i \), and phase \( \phi_i \) values for the \( i \)-th order can be determined by fitting the observed light-curve data using Eq. (\ref{eq:1}).

However, for Blazhko RRab stars, Eq. (\ref{eq:1}) cannot be directly applied to analyze the light curves due to amplitude and/or phase modulation, which is 'noisier' than that of Non-Blazhko RRab stars, as documented by \cite{2016ApJS..227...30N}. To infer the Blazhko frequencies of the stars from the power spectra, we adopt the methods outlined by \cite{2003ApJ...598..597A} and \cite{2013ApJ...773..181N}. Based on the approach described in the literature \citep{2012MNRAS.424.3094S,2016ApJS..227...30N}, the modulated light curves of RRab stars are fitted using the following expression:
\begin{equation}
\label{eq:2}
m(t) = F_0(t,f_0)+F_m(t,f_0,f_m),
\end{equation}
where F$_0$ is the same as that in Eq.\ref{eq:1}, 
\begin{equation}
\begin{aligned}
F_0(t,f_0) = m_0+\sum \limits_{i=1}^{n} A_i {\rm sin}(2\pi i f_0 t+\phi_{i}), \notag
\end{aligned}
\end{equation}
and F$_m$ includes the modulation components:

\begin{equation}
\begin{aligned}
F_m(t,f_0,f_m)  = & \sum \limits_{j=1}^{r}A_j^m {\rm sin}(2\pi j f_m t+\phi_{j}^{m}) \\
                  & +\sum \limits_{k=1}^{q}A_k^{-}{\rm sin}[2\pi (kf_0-f_m)t+\phi_{k}^{-}] \\
                  & +\sum \limits_{k=1}^{q}A_k^{+}{\rm sin}[2\pi (kf_0+f_m)t+\phi_{k}^{+}].\notag
\end{aligned}
\end{equation}
In Eq.\ref{eq:2}, the modulation frequency \( f_m \), which can be inferred from the power spectra of Blazhko RRab stars \citep{2000ApJ...542..257A,2003ApJ...598..597A,2013ApJ...773..181N}, is defined for each star as the reciprocal of the difference between the side peaks and the main pulsation frequency in the frequency spectrum \citep{2020MNRAS.494.1237S}, determined using SigSpec \citep{2012CoAst.163....3R}.
We adopted the equation \({|f^{sig}_{0\pm} - f_0|}\) = \({f_m} \) from \cite{2020MNRAS.494.1237S}, selecting the side peak associated with the fundamental frequency \( f_0 \), where \( f^{sig}_{0\pm} \) is the side peak nearest to \( f_0 \) with the highest significance. The modulation periods were calculated and listed in Column (12) of Tabel~\ref{tab:Blazhko}.
A series of combinations of Fourier order (n,r,q) are examined to find the best-fitting combinations for fitting the modulated light curves. These parameters (n,q,r) were selected such that the amplitude of each component exceeded the noise level by a factor of 5.6, with typical values of \(n\) ranging from 8 to 12, and \(r\) in the range of 4 to 6, and \(q\) ranging from 7 to 10. The light curves were fitted using the curve$\_$fit of Python Scipy library \citep{2024zndo..12522488G}. The light-curve parameters, or Fourier parameters, can be expressed in terms of A$_{i}$ and $\phi_{i}$:
\begin{equation}
\label{eq:3}
\phi_{ij} = j\phi_{i}-i\phi_{j}, R_{ij} = \frac{A_i}{A_j},
\end{equation}
where \( j = 1 \), \( i = 2 \) and \( 3 \) for the fundamental pulsation of RRLs, as suggested by \cite{2013ApJ...773..181N} and \cite{2013MNRAS.428.3034S}. 
The pulsation parameters of $Kepler$ Non-Blazhko RRLs analyzed here are taken from the literature \citep{2011MNRAS.417.1022N}. The total amplitudes and rise times of the light curves of these Non-Blazhko RRab stars are determined based on the method of \cite{2023ApJ...945...18Z}. For the total amplitudes and rise times of the light curves of the Blazhko RRab stars studied in this work, we follow the method adopted by \cite{2020MNRAS.494.1237S}. Those pulsation parameters of Blazhko and Non-Blazhko RRab stars are listed in Coloum (2)-(7) of Table~\ref{tab:Blazhko} and ~\ref{tab:Non-Blazhko}, respectively.

\subsection{The determination of metallicities}
We cross-matched the catalog of RRab stars studied in this work with the catalog of LRS of LAMOST DR12, resulting in a total of 147 Non-Blazhko and 111 Blazhko RRab stars matched. The metallicities [Fe/H] of those stars are estimated from those spectra using the $\Delta$S method \citep{1959ApJ...130..507P}, which is a traditional method to measure the metallicities of RRLs and adopted in the literature of \cite{2020ApJS..247...68L}. Those metallicities of those Blazhko and Non-Blazhko stars were listed in the Column (8) of Table~\ref{tab:Blazhko} and Table~\ref{tab:Non-Blazhko}, respectively.

\subsection{Absolute magnitudes and colours}

In this work, we adopted equation (21) from \cite{2024A&A...684A.176P} to determine the absolute magnitudes of the two groups of stars, using metallicities derived from Gaia DR3 and main periods obtained from light curves observed by \(Kepler/K2\), while ensuring data consistency by relying solely on Gaia DR3 metallicities and excluding higher-precision values from LAMOST or Keck to avoid potential biases from different data sources. The absolute magnitudes of the two groups stars were listed in Column (10) of Table~\ref{tab:Blazhko} and \ref{tab:Non-Blazhko}. When determining absolute magnitudes, it is important to note that they represent the average brightness over a full pulsation cycle. For Blazhko stars, this average magnitude is also influenced by the Blazhko cycle.  \cite{2014MNRAS.445.1584S} estimated the absolute magnitudes of both Blazhko and Non-Blazhko RRab stars for comparison using formula (2) from \cite{1998A&A...333..571J}, which is based on the main period \(P\), amplitude \(A_1\), and phase difference \(\phi_{31}\), without accounting for the Blazhko cycle. They stated that their study aimed to compare Blazhko and non-modulated stars rather than to precisely determine their parameters. Similarly, \cite{2017MNRAS.466.2602P} applied the same approach to determine the absolute magnitudes of Blazhko and Non-Blazhko RRab stars for comparison using formula (3) from \cite{2004ApJS..154..633C}, which relies on the main period and metallicity, also without considering the Blazhko cycle.


The metallicities were derived by cross-matching the catalogues of the two groups with the Gaia DR3 catalogue \citep{2023A&A...674A..28F,2023A&A...674A..27A} using the software TOPCAT \citep{2005ASPC..347...29T,2006ASPC..351..666T}. This process yielded metallicities and colours for 766 Blazhko and 950 Non-Blazhko RRab stars, respectively. However, it was found that some stars in both groups either lacked metallicity values or colours \(Bp-Rp\) after the cross-matching process. The absolute magnitudes and colours of those Blazhko and Non-Blazhko RRab stars are listed in Coloum (10)-(11) of Table~\ref{tab:Blazhko} and Table~\ref{tab:Non-Blazhko}, respectively.

\begin{table*}   
\centering
    \setlength\tabcolsep{8pt}
    \caption{The parameters of those Blazhko RRab stars studied in this work. Column (1) is the star ID, Column (2)-(7) are the pulsation parameters, Column (8)-(9) are the metallicities and their corresponding S/N obtained from LRS of LAMOST DR12, respectively. Column (10)-(11) are the absolute magnitudes and colours determined from the data of Gaia DR3, respectively.  \textbf{Column(12) is the Blazhko periods.}}\label{tab:Blazhko}
    \begin{tabular}{cccccccccccc}
    \hline
    \hline
ID	&	P	&	\(R_{21}\)	&	\(R_{31}\)	&	\(\phi_{21}\)	&	\(\phi_{31}\)	&	\(RT\)	& [Fe/H] & S/N& \(M_G\) & \(Bp-Rp\)&\(P_m\)\\
  EPIC  &	(day)	&		&		&		&		&	& & & (mag) & (mag)& (day)	\\
    (1) & (2) & (3)  & (4) & (5) & (6) & (7) & (8) &(9) &(10) & (11)& (12)\\
 \hline
212106005	&	0.5466	&	0.4834	&	0.3112	&	2.282	&	4.785	&	0.183	&	-1.85 & 98.15 & 0.21& 0.56&28.56\\
...	&	...	&	...	&	...	&	...	&	...	&	...	&	... & ... & ...	& ...&\\
...	&	...	&	...	&	...	&	...	&	...	&	...	&	... & ... & ...	& ...&\\
\hline
    \end{tabular}
\end{table*}

\begin{table*}   
\centering
    \setlength\tabcolsep{8pt}
    \caption{The parameters of those Non-Blazhko RRab stars studied in this work. Column (1) is the star ID, Column (2)-(7) are the pulsation parameters, Column (8)-(9) are the metallicities and their corresponding S/N obtained from LRS of LAMOST DR12, respectively. Column (10)-(11) are the absolute magnitudes and colours determined from the data of Gaia DR3, respectively. }\label{tab:Non-Blazhko}
    \begin{tabular}{ccccccccccc}
    \hline
    \hline
 ID	&	P	&	\(R_{21}\)	&	\(R_{31}\)	&	\(\phi_{21}\)	&	\(\phi_{31}\)	&	\(RT\)	& [Fe/H] & S/N &\textbf{\(M_G\)} & \textbf{\(Bp-Rp\)} \\
EPIC	&	(day)	&		&		&		&		&	&  & &\textbf{(mag)} & \textbf{(mag)}\\
    (1) & (2) & (3)  & (4) & (5) & (6) & (7) & (8) &(9) &(10)&(11)\\
 \hline
220489863	&	0.5977	&	0.531	&	0.369	&	2.298	&	4.895	&	0.139 &  -1.75 & 13.17 & 0.067 & 0.626	\\
...	&	...	&	...	&	...	&	...	&	...	&	...	&	... & ... & ...	& ...\\
...	&	...	&	...	&	...	&	...	&	...	&	...	&	... & ... & ...	& ...\\
\hline
    \end{tabular}
\end{table*}

\section{Results}
\subsection{The properties of Fourier composition coeﬀicients}
\begin{figure*}
    \centering
    \includegraphics[width=5.7in]{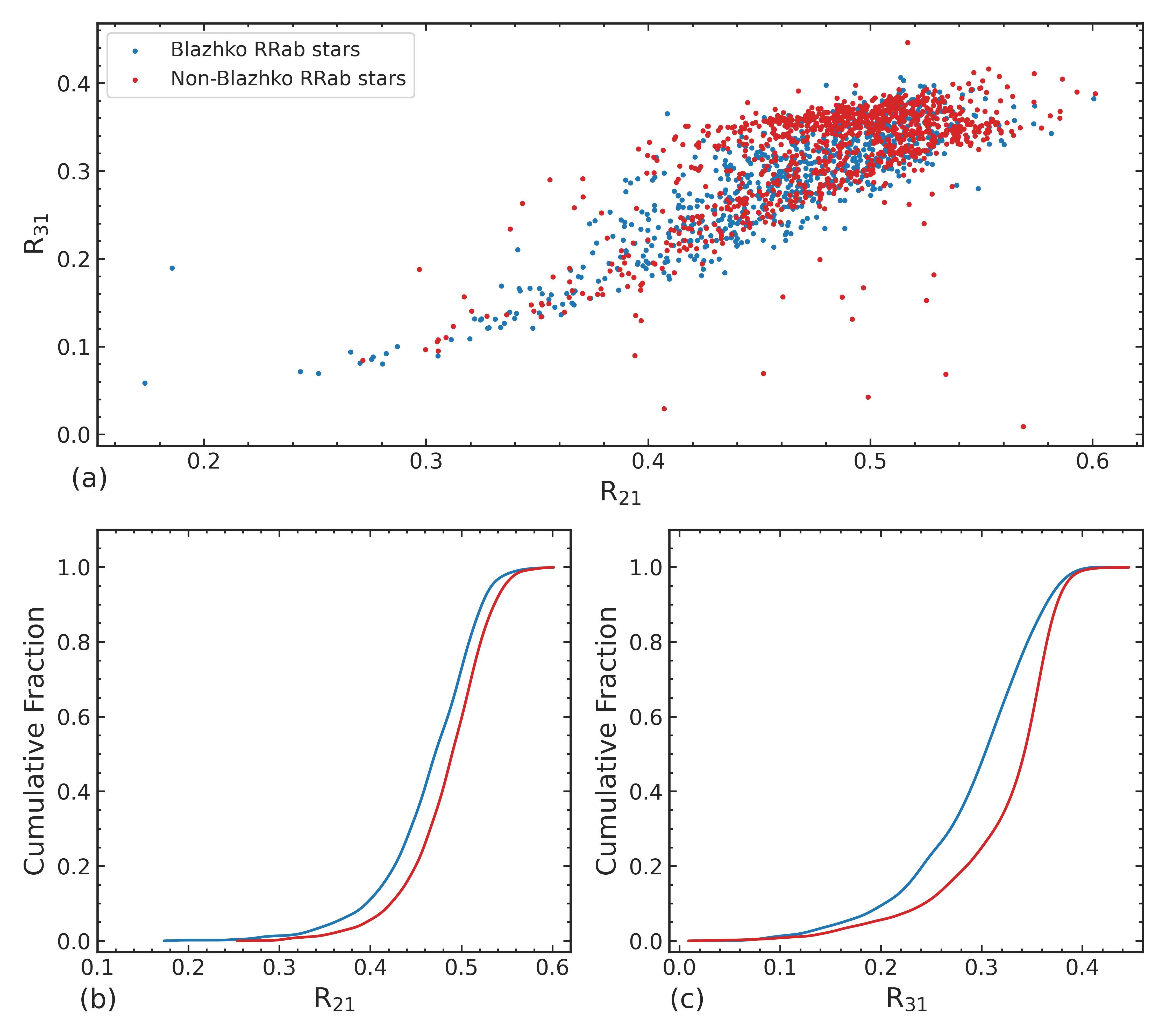}
    \caption{Fourier parameters of amplitude ratio diagram for those BL and nBL stars. Panel (a) shows the R$_{21}$ against R$_{31}$, panels (b) and (c) show the cumulative distribution function of the two parameters between the Blazhko and Non-Blazhko RRab stars with different colors.}
    \label{fig:Amp_ratio}
\end{figure*}

The low-order Fourier parameters of the phase differences \(\phi_{21}\) and \(\phi_{31}\), as well as the amplitude ratios R$_{21}$ and R$_{31}$, of the light curves can characterize the pulsation properties of RRLs \citep{1982ApJ...261..586S}.  \cite{2014MNRAS.445.1584S} conducted a comprehensive investigation to elucidate the correlations between \( R_{21} \) and \( R_{31} \) in Blazhko and Non-Blazhko RRab stars. We presented the analysis of the amplitude ratios \( R_{21} \) versus \( R_{31} \) of the Blazhko and Non-Blazhko RRab stars in this study. The distribution of amplitude ratios \( R_{21} \) versus \( R_{31} \) for both Blazhko and Non-Blazhko RRab stars is presented in panel (a) of Figure~\ref{fig:Amp_ratio} in different colors. We find that most Non-Blazhko RRab stars have \( R_{21} \) values ranging from approximately 0.36 to 0.56 and \( R_{31} \) values from approximately 0.33 to 0.40, while most Blazhko RRab stars show \( R_{21} \) values ranging from approximately 0.36 to 0.60 and \( R_{31} \) values from approximately 0.22 to 0.40. The standard deviations of the amplitude ratios \(R_{21}\) and \(R_{31}\) for Blazhko RRab stars are 0.054 and 0.064, respectively, while for Non-Blazhko RRab stars, they are 0.047 and 0.60, respectively. We tested the difference in the distributions of \( R_{21} \) and \( R_{31} \) for Blazhko and Non-Blazhko RRab stars using statistical tests of Kolmogorov–Smirnov (KS) test from the Python Scipy library \citep{2024zndo..12522488G}, as shown in panels (b) and (c), respectively. The p-value is then used to assess the significance of this difference, with values below 0.05 indicating a statistically significant result. The KS test yielded p-values for the amplitude ratios \( R_{21} \) and \( R_{31} \) of Blazhko and Non-Blazhko RRab stars are \(5.4 \times 10^{-14}\) and \(1.16 \times 10^{-38}\), respectively, which are less than the critical value of 5\% (0.05), indicating that significant differences exist between the two groups in terms of their amplitude ratios \( R_{21} \) and \( R_{31} \).

\begin{figure*}
    \centering
    \includegraphics[width=5.7in]{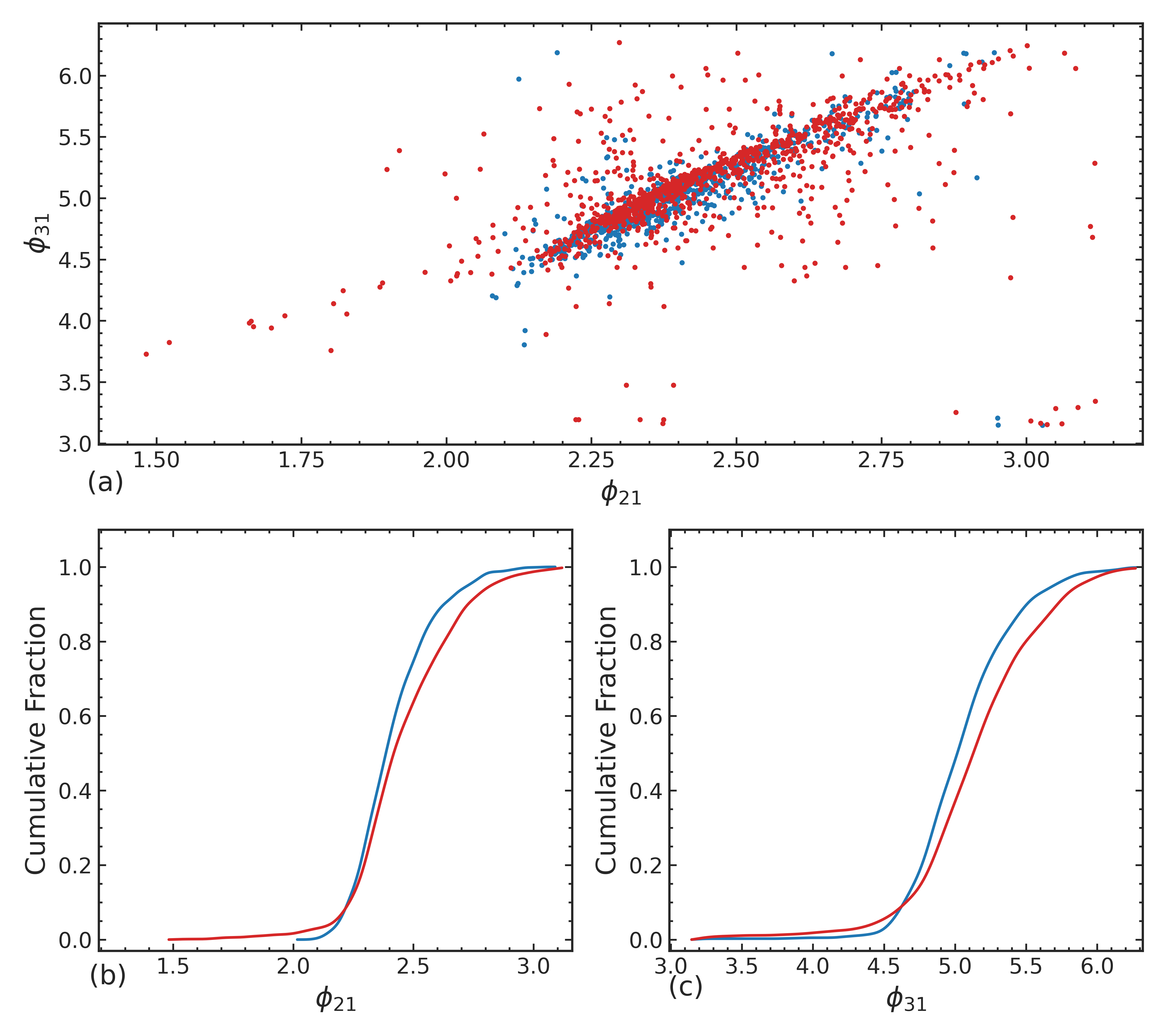}
    \caption{Fourier parameters of $\phi_{21}$ versus $\phi_{31}$ diagram for those BL and nBL stars. Panel (a) presents the relation between the two parameters. The cumulative distribution function of the two parameters  are shown in panel (b) and (c), respectively.}
    \label{fig:PHI}
\end{figure*}

The paper of \cite{2014MNRAS.445.1584S} and \cite{2017MNRAS.466.2602P} carried out comparison studies in phase differences of \(\phi_{21}\) and \(\phi_{31}\) determined from the light curves of Blazhko and Non-Blazhko RRab stars. In this work, we present a comparative analysis of these phase differences for the two types of stars, with the results depicted in Figure~\ref{fig:PHI}. Panel (a) of Figure~\ref{fig:PHI} shows the distribution of phase differences \(\phi_{21}\) versus \(\phi_{31}\) in different colors, a higher \(\phi_{21}\) is generally associated with a higher \(\phi_{31}\) for most Blazhko and Non-Blazhko RRab stars. The standard deviations of the phase differences \(\phi_{21}\) and \(\phi_{31}\) for Blazhko RRab stars are 0.16 and 0.36, respectively, while for Non-Blazhko RRab stars, they are 0.21 and 0.46, respectively. The KS test was used to assess the differences in the distributions of \(\phi_{21}\) and \(\phi_{31}\), as shown in panels (b) and (c) of this figure. The p-values for both phase differences \(\phi_{21}\) and \(\phi_{31}\) are \(7.26 \times 10^{-7}\) and \(3.47 \times 10^{-9}\), respectively, which are below the 0.05 threshold, showing that the phase differences \(\phi_{21}\) and \(\phi_{31}\) are statistically significant between the two groups.

\begin{figure*}
\begin{center}
\includegraphics[width=7.0in]{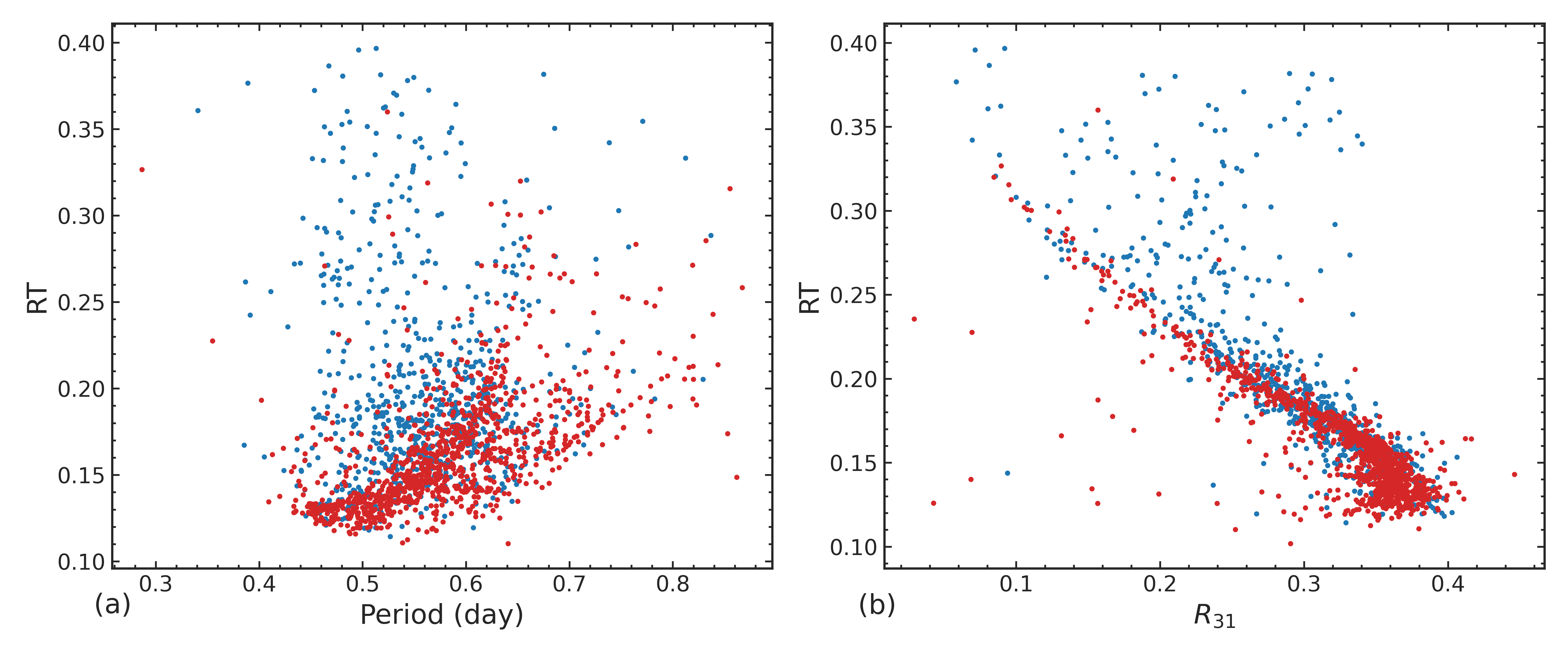}
\caption{Comparing the \(RT\) with other pulsation parameters of main periods with \(R_{31}\)for Non-Blazhko and Blazhko RRab stars. Panel (a) and (b) shows the \(Period\) vs \(RT\) and \(R_{31}\) vs \(RT\), respectively. }
\label{fg:RT_R31}
\end{center}
\end{figure*}

The rise times \(RTs\) of RRLs play a crucial role in shaping their light curves. In the study by \citet{2023ApJ...945...18Z}, the correlation between \(RTs\) and pulsation parameters of RRLs was investigated, but their study focused solely on Non-Blazhko RRab stars in the LAMOST-\textit{Kepler/K2} field and did not include Blazhko RRab stars. An earlier study by \citet{2014MNRAS.445.1584S} performed a comparative investigation of \(RTs\) with pulsation parameters for both Blazhko and Non-Blazhko RRab stars. Later, \citet{2017MNRAS.466.2602P} studied the correlation between the pulsation periods and \(RTs\) of the Blazhko and Non-Blazhko RRab stars obtained from the OGLE-IV photometric data. In this work, we carry out a comparative investigation of the rise times \(RTs\) with pulsation parameters of main periods and amplitude ratios \( R_{31} \) of the Blazhko and Non-Blazhko RRab stars, showing in Figure~\ref{fg:RT_R31}, with each group stars represented in different colors. The main periods against RTs for Blazhko and Non-Blazhko RRab stars are presented in panel (a) of this figure. The Pearson correlation coefficients \citep{2024zndo..12522488G} reveal that there is no significant correlation, as shown by the large (close to one) value of p (r = -0.006, p = 0.87) for Blazhko RRab stars, while a statistically significant moderate positive correlation (\( r = 0.54 \), \( p = 0.0 \)) is observed for Non-Blazhko RRab stars. The standard deviations of \(RTs\) further highlight these differences, with values of 0.058 for Blazhko and 0.035 for Non-Blazhko stars, respectively, suggesting that the \(RT\)s of Blazhko RRab stars are more scattered. Interestingly, the variability in main periods shows an inverse trend, with standard deviations of 0.66 for Blazhko and 0.80 for Non-Blazhko stars. The relationship between \( R_{31} \) and \(RT\) for Blazhko and non-Blazhko RRab stars is investigated by computing the Pearson correlation coefficient using the scipy library in Python \citep{2024zndo..12522488G}. The results show a strong negative correlation, with \( r = -0.78 \) (\( p = 0 \)) for Blazhko stars and \( r = -0.85 \) (\( p = 0 \)) for non-Blazhko RRab stars, indicating a significant inverse relationship between \( R_{31} \) and RT in both groups. The comparison of standard deviations for \(R_{31}\) and \(RTs\) reveals greater variability in Blazhko stars compared to Non-Blazhko stars. This suggests that the distribution of \(R_{31}\) versus \(RT\) for Non-Blazhko RRab stars is more scattered than that of Blazhko RRab stars.

\subsection{Metallicities and absolute magnitudes with colours of the two types stars}
\begin{figure*}
    \centering
    \includegraphics[width = 5.7 in]{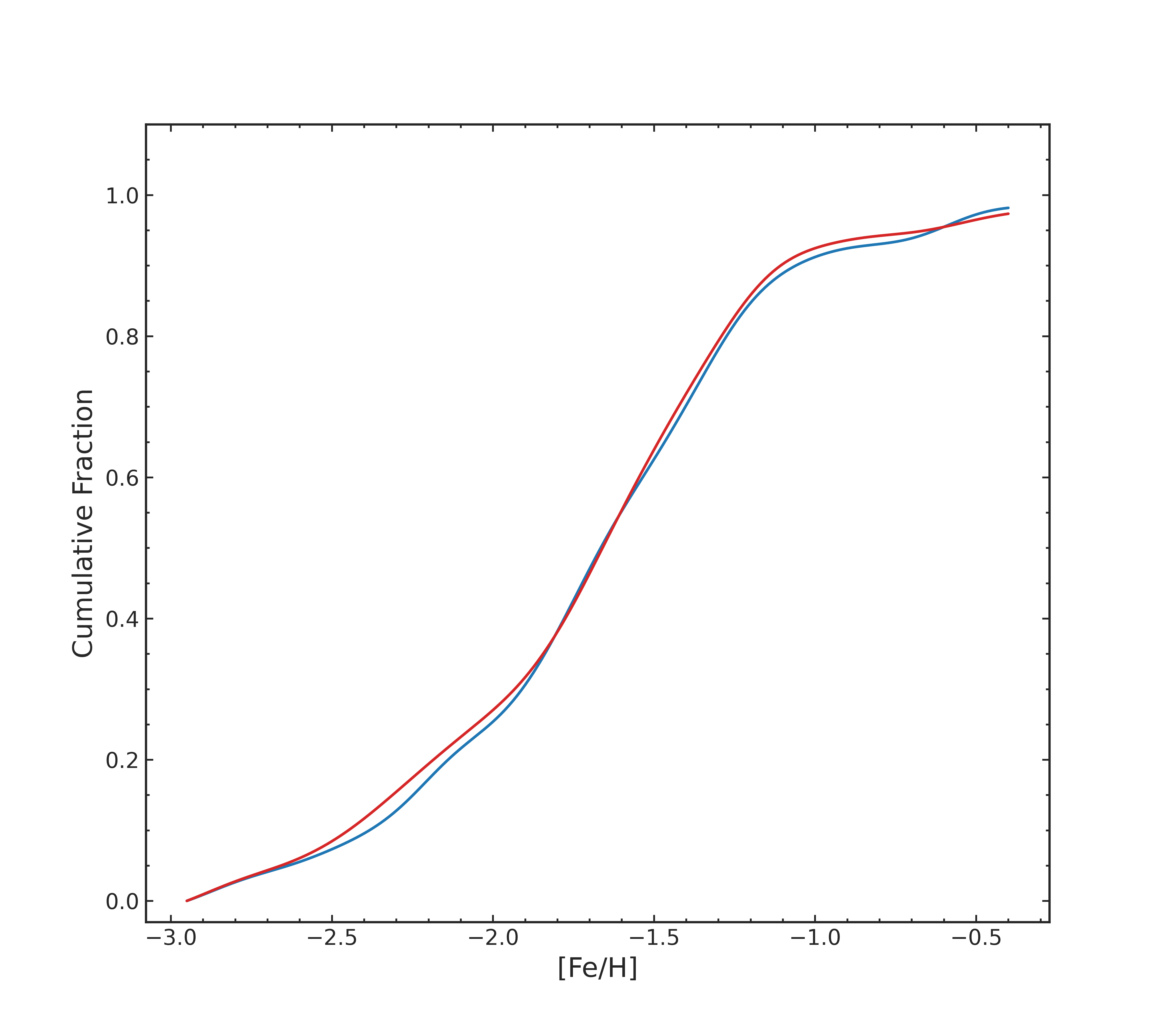}
    \caption{The cumulative distribution function of the two parameters of the metallicities of Blazhko and Non-Blazhko stars.}
    \label{fig:FeH}
\end{figure*}

As documented by \cite{2014MNRAS.445.1584S}, metallicity is one of the most critical parameters for comparing the differences between Blazhko and Non-Blazhko RRLs. The metallicity differences for Blazhko and Non-Blazhko RRab stars have been studied in the literature \citep{2003A&A...398..213M,2005AcA....55...59S,2013ApJ...773..181N,2014MNRAS.445.1584S}. In this work, we use the metallicities of RRab stars determined from the low-resolution spectral surveys of LAMOST DR12 to investigate the difference in metallicities between Blazhko and Non-Blazhko RRab stars. The results are presented in Figure~\ref{fig:FeH}. The KS test was applied to compare the metallicity distributions of Blazhko and Non-Blazhko RRab stars, as illustrated in Figure~\ref{fig:FeH}. The p-value of 0.99 is well above the 0.05 significance threshold, indicating no statistically significant difference between the metallicity distributions of the two groups.

\begin{figure*}
    \centering
    \includegraphics[width = 5.7 in]{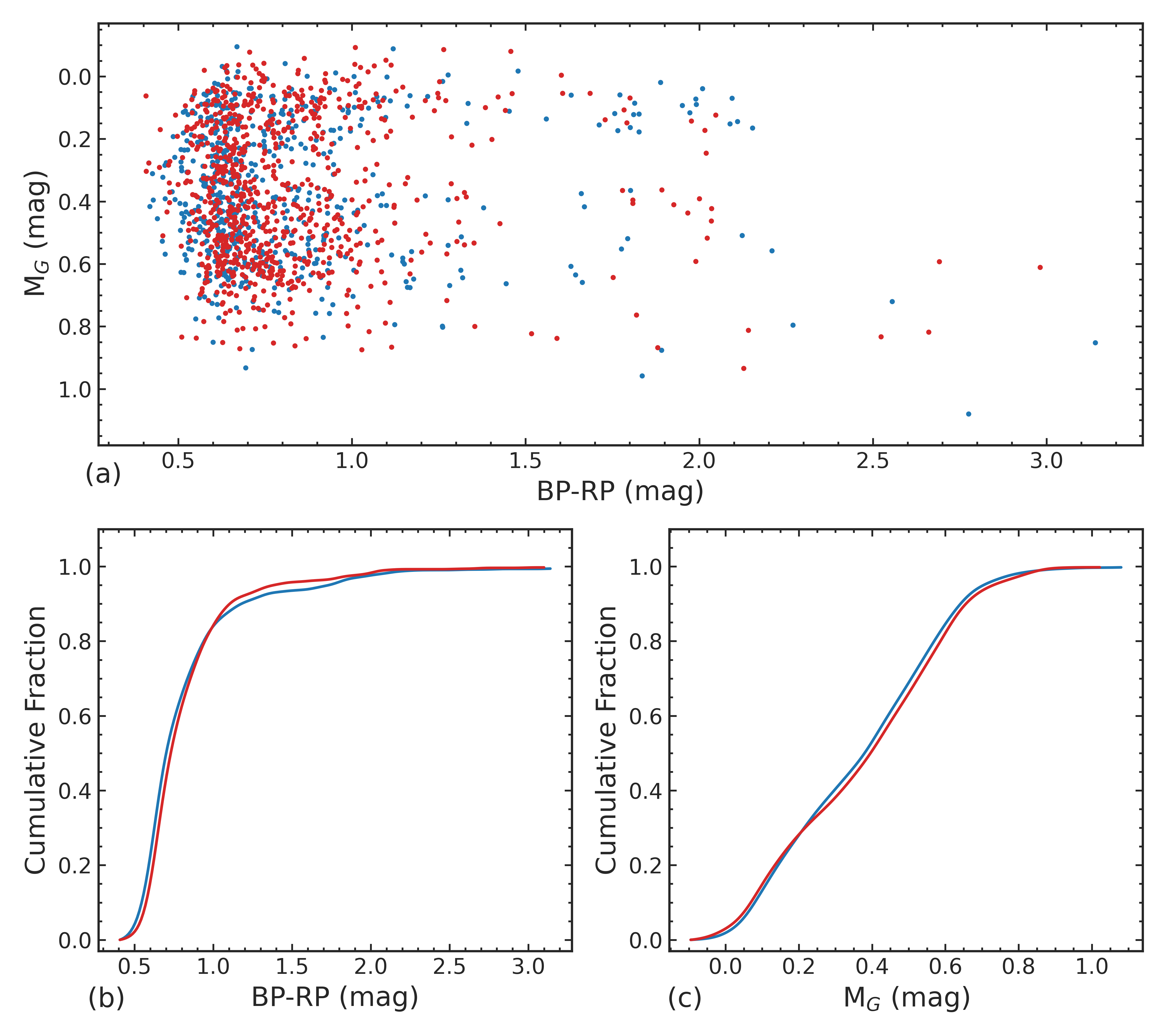}
    \caption{The absolute magnitudes and colour indices of Blazhko and Non-Blazhko RRab stars. Panel (a) presents the distributions of $colours$ vs absolute magnitudes. The cumulative distribution function of the two parameters are shown in panel (b) and (c), respectively.}
    \label{fig:Comparing}
\end{figure*}

Panel (a) of Figure~\ref{fig:Comparing} displays the absolute magnitudes against colors, calculated using the formula (21) of \citep{2024A&A...684A.176P} for the Blazhko and Non-Blazhko RRab stars analyzed in this work. To compare the distributions of absolute magnitudes and color indices for Blazhko and Non-Blazhko RRab stars, we conducted the KS test, as shown in panels (b) and (c) of Figure~\ref{fig:Comparing}. The p-value for absolute magnitudes is 0.61, which is significantly greater than the 0.05 threshold, while the p-value for color indices is 0.0004, which falls below the threshold. This suggests that the distributions of absolute magnitudes for the two groups are statistically indistinguishable, whereas their color index distributions exhibit a statistically significant difference.

\section{Discussion}
\subsection{Comparing with RRab stars in LMC field}
\begin{figure*}
\begin{center}
\includegraphics[width=7.0 in]{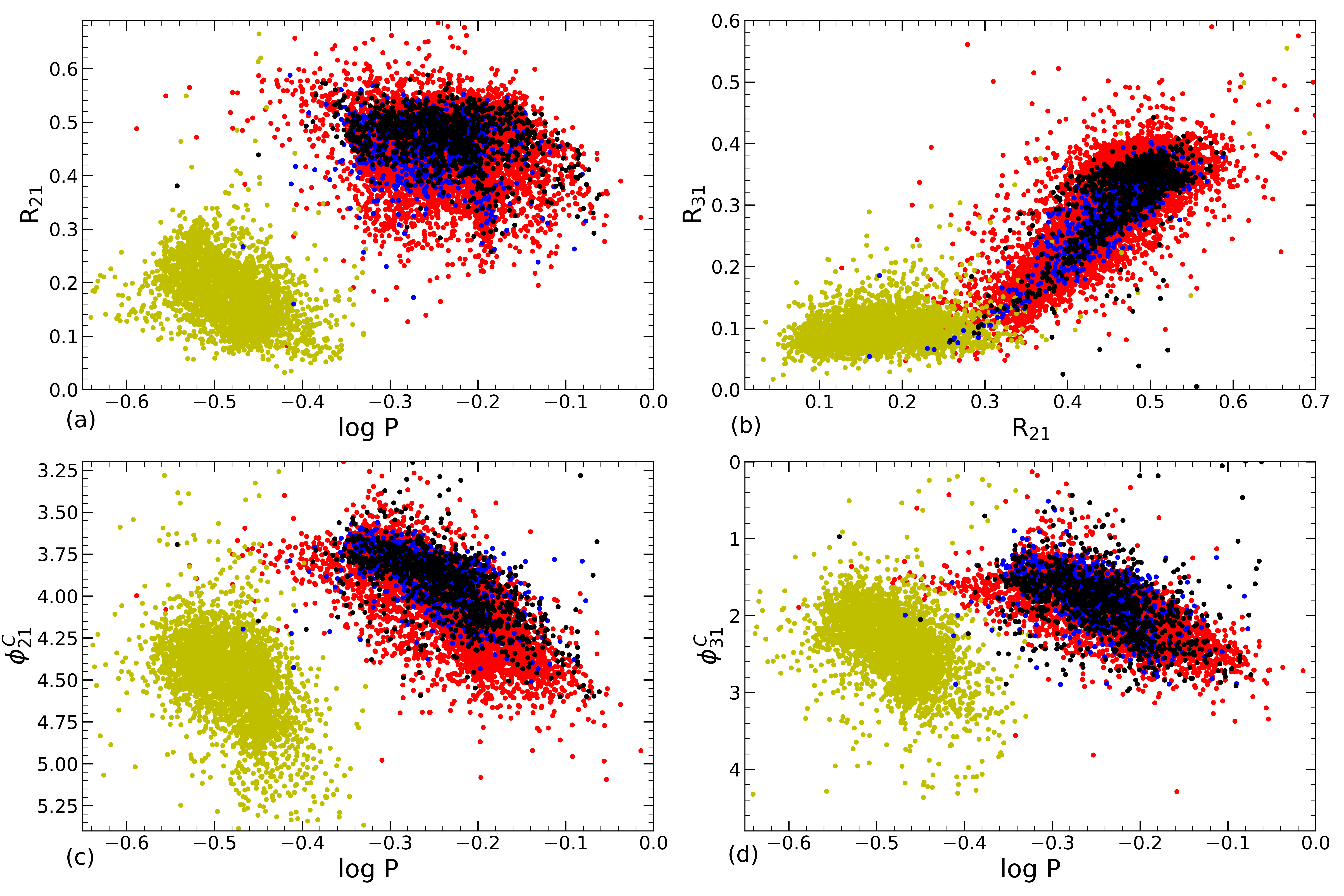}
\caption{Comparing of Fourier parameters with OGLE-IV LMC field RR Lyrae stars \citep{2016AcA....66..131S}. The red points represent RRab, yellow points RRc stars. The black and blue pentagon are Non-Blazhko and Blazhko stars in the field of \(Kepler/K2\). }
\label{fg:Com_LMC}
\end{center}
\end{figure*}

The Fourier decomposition coefficients \(R_{21}\), \(R_{31}\), \(\phi_{21}\), and \(\phi_{31}\) of the Blazhko and Non-Blazhko RRab stars studied in this work are transformed from the \(K_p\) band to the \(V\) band using the equations provided by \cite{2011MNRAS.411.1763J}. We compare these Fourier decomposition coefficients with those derived for RRLs in the central regions of the LMC, determined from the OGLE Collection of Variable Stars by \cite{2016AcA....66..131S} and transformed from the \(I\)-band to the \(V\)-band using equations provided by \cite{1998AcA....48..341M}. The comparison results are presented in Figure~\ref{fg:Com_LMC}. It is clear that the coefficients of the Blazhko and Non-Blazhko RRab stars determined in this work align well with those of the RRab stars and differ from those of other types of RRLs shown in Figure~\ref{fg:Com_LMC}. Note that a few Blazhko RRab stars lie near the crossover between RRab and RRc stars, as shown in the top-right panel of the figure. However, detailed investigations of these stars are beyond the scope of this work.

\subsection{The properties of pulsation parameters}
\begin{figure*}
\begin{center}
\includegraphics[width=7.0 in]{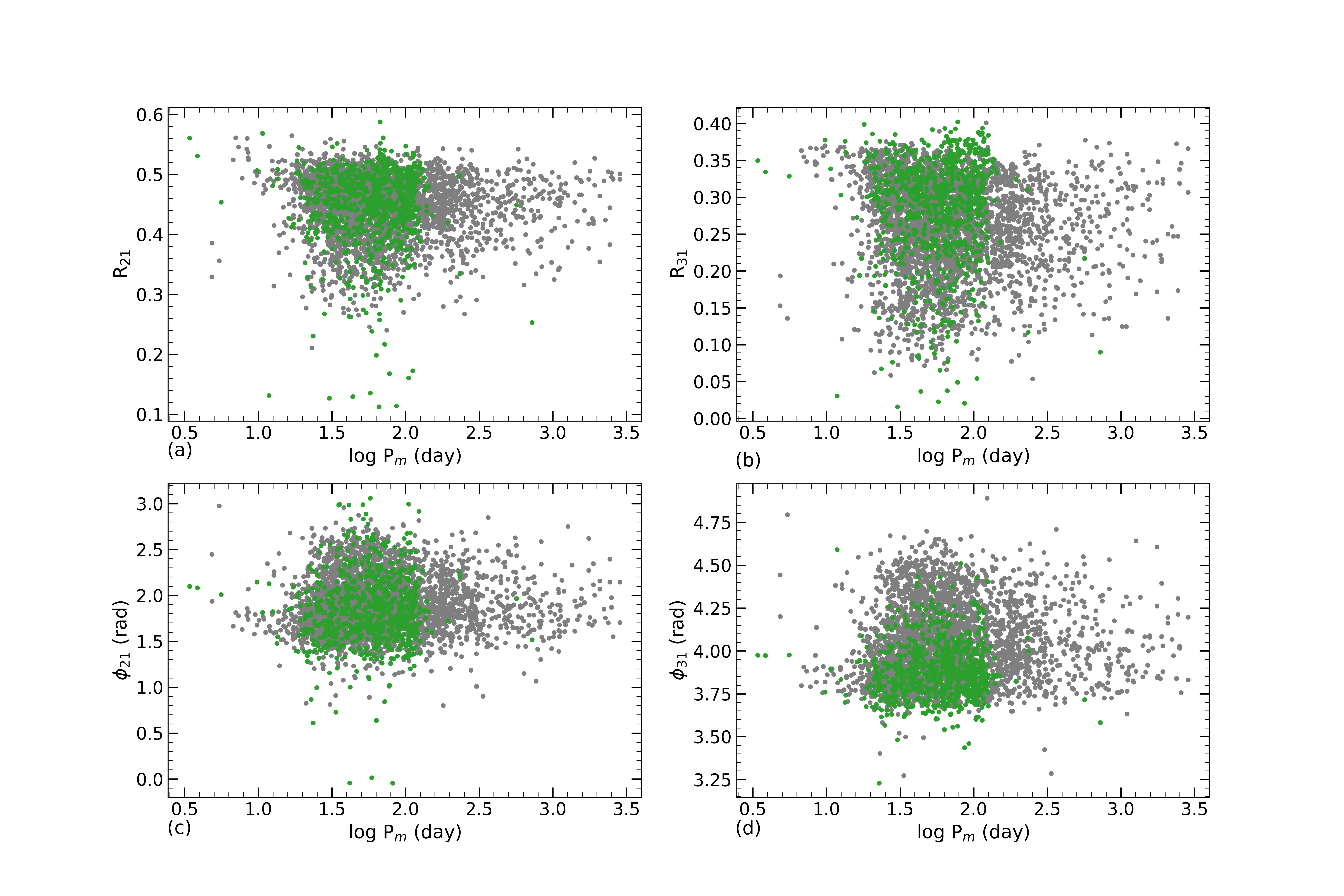}
\caption{Low-degree Fourier coefficients of the sample stars. Amplitude coefficients \(R_{21}\), \(R_{31}\) (top panels) and phase coefficients \(\phi_{21}\) and \(\phi_{31}\) (bottom panels), as a function of the Blazhko period are shown. The gray points are the Blezhko RRab stars from the study of \cite{2020MNRAS.494.1237S}, the green points are the Blazhko RRab stars determined in this work.}
\label{fg:Blazhko_Comparison}
\end{center}
\end{figure*}

Table~\ref{tab:mean} lists the mean values of the pulsation parameters of the Blazhko and Non-Blazhko RRab stars studied in this work. The uncertainties of those parameters are the standard error of 1 $\sigma$ of those values. The mean values of the pulsation periods, the Fourier parameters amplitude ratios and phase differences of Blazhko RRab stars are within the uncertainties of those values of Non-Blazhko RRab stars. This is different from that result of \cite{2017MNRAS.466.2602P}, who found that the mean values of the Fourier parameters are systematically lower for Blazhko stars than for Non-Blazhko stars as listed in their Table 2. 
In this study, we find the statistically difference in the distributions of the amplitude ratios \( R_{21} \) and \( R_{31} \) determined from the light curves of Blazhko and Non-Blazhko RRab stars observed in the fields of \(Kepler/K2\). The same phenomenon is also found in the distributions of the phase differences \(\phi_{21} \) and \( \phi_{31} \) for these Blazhko and Non-Blazhko RRab stars. However, distinguishing between Blazhko and Non-Blazhko RRab stars based on Fourier parameters remains challenging, as the two groups exhibit significant overlap, as illustrated in panels (a) of Figures~\ref{fig:Amp_ratio} and \ref{fig:PHI}.Interestingly, the standard deviation \(\phi_{21}\) is smaller in Blazhko RRab stars (0.17) than in Non-Blazhko RRab stars (0.22), and similarly for \(\phi_{31}\) (0.37 vs. 0.46). The standard deviation of \(\phi_{31}\) is significantly larger than that of \(\phi_{21}\) for both Blazhko and Non-Blazhko RRab stars, which implies that the distribution of the higher-order phase difference \(\phi_{31}\) is more dispersed than that of the lower-order phase difference \(\phi_{21}\). As shown in Figure~\ref{fig:Amp_ratio}, Blazhko RRab stars are primarily distributed along the stem, while 54\% of Non-Blazhko RRab stars are concentrated in the pinhead region where \( R_{31} > 0.34 \).  Approximately 48\% of the stars are Blazhko stars at \( R_{31} < 0.3 \), with 11.2\% being stable stars at \( R_{31} < 0.25 \). Our results differ from those reported in the literature \citep{2014A&A...562A..90S}, which may be attributed to the larger sample size of our study and the higher precision of the photometric data compared to those used in their work.Figure~\ref{fg:Blazhko_Comparison} shows the distributions between the amplitude ratios with phase differences and the length of the modulation periods of the Blazhko RRab stars.The relationships between modulation periods and pulsation parameters, specifically amplitude ratios and phase differences, in Blazhko RRab stars were investigated through the computation of Pearson correlation coefficients using the scipy library in Python \citep{2024zndo..12522488G}. The results of this analysis revealed the p-values of Pearson correlation coefficients are 0.13 (log P$_m$ vs \(R_{21}\)), 0.57 (log P$_m$ vs \(R_{31}\)), 0.21 (log P$_m$ vs \(\phi_{21}\)) and 0.25 (log P$_m$ vs \(\phi_{31}\)), suggesting that no  statistically significant correlations between the modulation periods and the pulsation parameters,as all p-values exceeded 0.05, which is consisted with the result that derived by \cite{2020MNRAS.494.1237S}.

For the rise times (\(RT\)) of the two types of stars determined in this work, the mean value of \(RT\) for the Blazhko RRab stars is 0.19 (6), while for the Non-Blazhko RRab stars, it is 0.16 (2), which falls within the uncertainties of the Blazhko RRab stars' \(RT\) values. This result differs from that of \cite{2017MNRAS.466.2602P}, who found a significant difference in the mean \(RT\) values between the two groups. This discrepancy may be due to the greater accuracy photometric data used in our study compared to those employed by \cite{2017MNRAS.466.2602P}. They observed that the \(RT\) values for Blazhko RRab stars were highly scattered, with most of them exhibiting higher \(RT\) values compared to the Non-Blazhko RRab stars. This finding is consistent with our results, as shown in Figure~\ref{fg:RT_R31}. We find that the \(RT\) values for Non-Blazhko RRab stars do not follow a clear linear trend, as most Non-Blazhko stars have rise times concentrated between approximately 0.1 and 0.16, as shown in panel (b) of Figure~\ref{fg:RT_R31}. This differs from the results of \cite{2014MNRAS.445.1584S}, and the discrepancy may be due to the larger sample size and higher precision of the photometric data in our analysis. However, no astrophysical interpretations have been offered in the literature regarding the differences in \(RT\) values between Blazhko and Non-Blazhko RRab stars.

\begin{table*}
    \centering
    \setlength\tabcolsep{13pt}
    \caption{Comparison of the mean parameters of Non-Blazhko and Blazhko RRab stars. The uncertainties in the last digits (in parentheses) are the standard 1$\sigma$ errors of the means. Fourier parameters are based on a sine fit.}\label{tab:mean}
    \begin{tabular}{ccccccc}
    \hline
     Type  &P$_{puls}$ (day)  & R$_{21}$ & R$_{31}$ & $\phi_{21}$ (rad) & $\phi_{31}$ (rad) & RT (rad) \\
   \hline
     nBL     & 0.58 (8)  & 0.48 (4) & 0.32 (5) &   2.45 (21) & 5.13 (46) & 0.16 (4)  \\
      BL     & 0.56 (6) & 0.46 (5) & 0.29 (6) &   2.41 (16) & 5.04 (36) & 0.20 (6)      \\
    \hline
    \end{tabular}
\end{table*}

\subsection{Metallicities and absolute magnitudes with colours}
In this work, we compare the metallicities derived from low-resolution spectra observed by LAMOST DR12 to investigate the properties of Blazhko and Non-Blazhko RRab stars. The KS test result indicates no difference in metallicities between the Blazhko and Non-Blazhko RRab stars. Our result is consistent with that of \cite{2017MNRAS.466.2602P}, who pointed out that the presence of the Blazhko effect in a particular star depends more on other physical parameters than on the metal abundance. \cite{2014MNRAS.445.1584S} performed the analysis of the differences in metallicities between Blazhko and Non-Blazhko RRab stars. They found weak indications that Blazhko stars could have slightly lower metallicities than Non-Blazhko stars, but no convincing proof was found. We find that their metallicities were estimated using the calibration from \cite{1996A&A...312..111J}, who derived that calibration using photometric data collected from heterogeneous observations at various sites and often lacked adequate phase coverage or had excessive noise from 81 RRab field stars in the \(V\) band with metallicities determined from the high-dispersion spectroscopy scale of \cite{1995AcA....45..653J}. \cite{2013ApJ...773..181N} analyzed metallicities estimated from the spectra observed by the Keck telescope and CFHT of Kepler field Blazhko and Non-Blazhko RRab stars. They found that the metallicities of 5 Non-Blazhko RRab stars were more metal-rich than [Fe/H] $\sim$ -0.9 dex, while all of the 16 Blazhko stars were more metal-poor. This contrasts with our results, as presented in Figure~\ref{fig:FeH}, which show that the distributions of metallicities for Blazhko and Non-Blazhko RRab stars are similar. This might be due to the fact that our sample stars have metallicities that are much larger than those of \cite{2013ApJ...773..181N}.

In this study, we find that there is no difference in absolute magnitudes between the two groups, which is not consistent with the findings in the literature \citep{2011MNRAS.411.1763J}. Based on the directly observed properties of RRLs in M5, \cite{2011MNRAS.411.1763J} found a difference in absolute magnitude between Blazhko and Non-Blazhko RRab stars. However, \cite{2014MNRAS.445.1584S} suggested that the limited sample size of RRLs in M5 alone could be misleading. \cite{2014MNRAS.445.1584S} also found a difference in absolute magnitude between the Blazhko and Non-Blazhko RRab stars, and they pointed out that lower luminosity may be a general property of Blazhko variables, which is different from the result derived in this work. This might be caused by the fact that their absolute magnitudes of the stars studied in their work were calculated using equation (2) from \cite{1998A&A...333..571J}. We also note that the absolute magnitudes of Blazhko and Non-Blazhko RRab stars determined in this study were derived using equation (21) from \cite{2024A&A...684A.176P}, which utilizes the primary periods obtained from light curves observed in the \(Kepler\)/\(K2\) field and the metallicities provided by Gaia DR3. However, recent studies \citep{2023A&A...674A..29R,2023A&A...676A.140O,2024MNRAS.533..395L} suggest that Gaia's capability for stellar abundance estimation may be somewhat limited. Additionally, the influence of Blazhko cycles was not accounted for in the absolute magnitude calculations, which may contribute to the discrepancies between our results and those of previous studies. Regarding the colors, our results are inconsistent with those of \cite{2014MNRAS.445.1584S} and \cite{2012MNRAS.420.1333A}, who found no apparent difference in colors between Blazhko and Non-Blazhko RRab stars. This discrepancy may arise from the fact that our sample size is significantly larger than theirs. The limited number of sample stars used in their studies, particularly within M5, could potentially lead to misleading conclusions, as noted by \cite{2014MNRAS.445.1584S}.

\section{Conclusion}
In this work, we carried out a comparison study in the properties of the Blazhko and Non-Blazhko RRab stars. We identified 1054 non-Blazhko and 785 Blazhko RRab stars in the photometric data observed by K2 mission, which, combined with those 37 stars observed in the original Kepler field, constituted our study sample. The periods and pulsation parameters were determined by applying the Fourier Decomposition method to the light curves of those stars. The pulsation parameters of amplitude ratios \(R_{21}\), \(R_{31}\) and phase differences \(\phi_{21}\), \(\phi_{31}\) are consisted with those parameters of RRab stars in the LMC. The pulsation parameters of amplitude ratios and phase differences of the Non-Blazhko stars are different from those of Blazhko stars. However, distinguishing between the two groups based on Fourier parameters was not feasible due to significant overlap in their distributions. The pulsation parameters of amplitude ratios and phase differences have no realtions with the modulation periods of Blazhko RRab stars, which is consisted with that in the literature. As for the \(RTs\) of those stars, we found that the values of RTs of the Blazhko RRab stars are very scattered and most of them have higher values than that values of RT s of the Non-Blazhko RRab stars.
  
By cross-matching with LRS of LAMOST DR12, we have obtained 147 Non-Blazhko and 111 Blazhko RRab stars of our sample stars observed in the LAMOST-\(Kepler/K2\) field. However, the distributions of the metallicities of Blazhko and Non-Blazhko stars have no difference, which is not consisted with that result in previous study. The absolute magnitudes and colours of 766 Blazhko and 950 Non-Blazhko RRab stars were determined in this work by cross-matching our sample stars with the Gaia DR3 catalog. Our analysis revealed differences in colours but no notable differences in absolute magnitudes, which is not consisted with that in the literature. Our findings indicate that the Blazhko effect of RRab stars may be more closely related to pulsation parameters, absolute magnitudes, or colors, rather than metallicities. In the future, a larger amount of precise light curves and spectra of RRLs will hopefully be obtained, which would bring new insight of the Blazhko effect of RRLs and help improve our understanding of the stellar evolution of RRLs.

\begin{acknowledgments}
We acknowledge the support from the National Natural Science Foundation of China (NSFC) through grants No. 12090040, 12090042, and 11833002. The Guoshoujing Telescope (the Large Sky Area Multi-object Fiber Spectroscopic Telescope, LAMOST) is a National Major Scientific Project built by the Chinese Academy of Sciences. The authors gratefully acknowledge the Kepler team and all who have contributed to making this mission possible.
Bo Zhang acknowledges the support from National Natural Science Foundation of China (NSFC) under grant No.12203068. The authors would like to extend their sincere thanks to the reviewer for the valuable feedback and constructive suggestions provided during the review process. The reviewer's insightful comments have played a significant role in enhancing the quality, clarity, and scientific rigor of this manuscript. We deeply appreciate the time and effort dedicated to improving our work.
\end{acknowledgments}

\software{astropy \citep{2013A&A...558A..33A, 2018zndo...4080996T},  
          LightKurve \citep{2018zndo...1412743B, 2021zndo...1181928B},
          Period04 \citep{2004IAUS..224..786L,2005CoAst.146...53L}
          }




\bibliography{sample631}{}
\bibliographystyle{aasjournal}

\end{document}